\documentclass[12pt]{iopart}

\usepackage{graphicx}
\usepackage{bm}
\usepackage{iopams}

\begin{document}


\title{Phase separation near the charge neutrality point in FeSe$_{1-x}$Te$_{x}$ crystals with x $<$ 0.15}

\author{Y.A.~Ovchenkov$^{1*}$, D.A.~Chareev$^{2,3.4}$, E.S.~Kozlyakova$^{1,4}$,  E.E.~Levin$^{5,6}$, M.G.~Miheev$^1$, D.E.~Presnov$^{1,7,8}$, A.S.~Trifonov$^{1,7}$, O.S.~Volkova$^{1,3,4}$ and A.N.~Vasiliev$^{1,3,4}$}

\address{$^1$Faculty of Physics, M.V. Lomonosov Moscow State University, Moscow 119991, Russia}

\address{$^2$Institute of Experimental Mineralogy (IEM RAS), Chernogolovka 142432, Russia}
\address{$^3$Ural Federal University,  Ekaterinburg 620002, Russia}
\address{$^4$National University of Science and Technology  'MISiS', Moscow 119049, Russia}
\address{$^5$ Faculty of Chemistry, M.V. Lomonosov Moscow State University, Moscow 119991, Russia}
\address{$^6$FSRC “Crystallography and Photonics” RAS, Moscow 119333, Russia}
\address{$^7$MSU Quantum Technology Centre, Moscow 119991, Russia}
\address{$^8$Skobeltsyn Institute of Nuclear Physics, Moscow 119991, Russia}

\ead{$^*$ovtchenkov@mig.phys.msu.ru}

\begin{abstract}
Our study of FeSe$ _ {1-x}$Te$ _ {x}$ crystals with x $<$ 0.15 shows that the phase separation in these compositions occurs into phases with a different stoichiometry of iron. This phase separation may indicate structural instability of the iron plane in the studied range of compositions.
To explain it, we discuss the bond polarity and the peculiarity of the direct $d$ exchange in the iron plane in the framework of the basic phenomenological description such as the Bethe-Slater curve. With this approach, when the distance between iron atoms is close to the value at which the sign of the magnetic exchange for some $d$ orbitals changes, structural and electronic instability can occur. Anomalies in the crystal field near the point of charge neutrality can also be a significant component of this instability. A similar instability of the iron plane may also be an important factor for other series of iron-based superconductors.

\end{abstract}

\maketitle


\section{Introduction}

Crystallographic phase instability is an important factor for enhancing superconductivity \cite{matthias1973criteria}. Some anomalies in the properties of compositions FeSe$ _ {1-x}$Te$ _ {x}$ with a low tellurium content, including the observation of phase separation in these compositions, indicate that phase instability may exist in these compounds.

The tetragonal plane of iron atoms, surrounded by pnictogen or chalcogen atoms, is the main motif of the crystal structure of iron-based superconductors (IBS). Therefore, the properties of this plane is of key importance for superconductivity in these compounds. In particular, the temperature of superconducting transitions depends on the structural parameters of this plane, such as, for example, the degree of deformation of the tetragonal environment of iron \cite{lee2008effect} or the distance from pnictogen to the iron plane \cite{kuroki2009pnictogen}.

The properties of the tetragonal plane of iron, surrounded by the pnictogens, could be thoroughly studied using almost ideal compositions of the 11 series of IBS \cite{hsu2008superconductivity,2017_Coldea, Bohmer2018}. The structure of these compounds is close to stoichiometric, and they have no intercalating elements that could distort the electronic properties of the iron plane. Nevertheless, to date, the properties of the 11 series of IBS have been studied only partially. There remains a whole range of compositions,  the properties of which have not yet been sufficiently studied. In particular, the synthesis of Fe(Se,Te) compounds with low tellurium content usually leads to phase separation in both crystals and films \cite{zhuang2014}. The nature of this phase separation is not yet been understood.

We studied the synthesis and properties of FeSe$ _ {1-x}$Te$ _ {x}$ crystals in the range of $x<0.15$. We find that for the studied compositions the phase separation manifests itself in transport properties as a distortion of the anomaly on $dR/dT$ curves near the structural transition points. For as-synthesized samples with different iron contents, the anomaly on $dR/dT$ splits into two distinct anomalies indicating the formation of two phases. Heat treatments of studied compositions can suppress phase separation, which leads to the disappearance of the anomaly in the $dR/dT$ curve located at a lower temperature. We also found suppression of phase separation in compositions after long-term storage, which suggests that, during annealing, the partial oxidation and removal of the excess of iron may have a major impact. Thus, the results obtained indicate that phase separation in the studied compositions is caused by deviations in the stoichiometry of iron. In turn, this may mean that phase separation in the studied compositions is a consequence of the structural instability of the iron plane at certain lattice parameters

One of the possibilities for describing the dependence of the properties of compounds with transition elements on their structure is to study the properties of individual orbitals or orbital-selective effects \cite{streltsov2017orbital}. The properties of an individual orbital state has a relatively universal dependence on the interatomic distance \cite{slater1930cohesion}, which for the case of direct $d$ interaction can be expressed by the Bethe-Slater curve. In FeSe, similar to pure metallic iron, the distance between the iron atoms is close to the value at which the sign of the direct magnetic exchange changes. Recent studies of cubic lattices of the elemental $3d$ metals have shown that the exchange interaction for Fe has different signs for different groups of orbitals \cite{cardias2017bethe}. Thus, the change in the sign of the interaction or, in other words, the degeneration of the singlet and triplet states can occur independently for different groups of the iron orbitals.

The electronic instability in series 11 for compositions close to FeSe has many experimental demonstrations  \cite{PhysRevB.96.100504, PhysRevB.95.224507}. For these compositions, small changes in the lattice parameters can produce significant changes in the electronic properties. In particular, a significant change in the mobility of carriers occurs in FeSe$ _ {1-x}$Te$ _ {x}$ with a low tellurium content  \cite{ovchenkov2019nematic, PhysRevB.100.224516}, which corresponds to a crossover from bad to good metal. This crossover accompanies the majority carriers inversion  in series 11. Formally, from neutrality, a change in the sign of mobile charges means an inversion of the charge of the ionic core. Thus, the inversion of carriers under an isovalent substitution \cite{Ovchenkov_2019} can be considered as evidence of a change in the polarity of the bond in these compounds, which in turn can be caused by the transformation of the $d$ orbitals with a change in the iron-iron distance in the plane.

The assumed structural instability of the iron/chalcogen plane near the charge neutrality point deserves further investigation as a possible important ingredient of superconductivity in IBS. Besides, this instability can be the reason for the splitting of some structural transitions in other series of iron-based superconductors.

\section{Experiment}

The studied crystals of FeSe${}_{1-x}$Te${}_{x}$ were prepared using the AlCl${}_{3}$/KCl/NaCl eutectic mixture in evacuated quartz ampoules in permanent gradient of temperature \cite{CrystEngComm12.1989, Char_FeSeS_CEC}. The quartz ampoules with the Fe(Te,Se) charge and maximum quantity of AlCl${}_{3}$/KCl/NaCl eutectic mixture were placed in a furnace so as to maintain their hot end at a temperature of 500~$^{\circ}$C and the cold end at a temperature of 433~$^{\circ}$C for $x = 0.15$ and $x=0.11$, and hot end at a temperature of 453~$^{\circ}$C and the cold end at a temperature of 400~$^{\circ}$C in the case of $x = 0.055$ . The chalcogenide charge is gradually dissolved in the hot end of the ampoule and precipitates in the form of single crystals at the cold end. After keeping for 8 weeks in the furnace, iron monochalcogenide platelike crystals were found at the cold ends of the ampoules.

To study the effect of heat treatment on the properties of crystals, two evacuated quartz ampoules with the $x=0.055$ crystals were heat-treated at the temperature 445~$^{\circ}$C for one week. Next, one of the ampoules was quenched in water, and another ampoule cooled with the oven. Other thermal treatments of crystals were also always carried out in evacuated ampoules.

The chemical composition of the crystals was determined using a Tescan Vega II XMU scanning electron microscope equipped with an INCA Energy 450 energy-dispersive spectrometer; the accelerating voltage was 20 kV.
For X-ray diffraction (XRD) analysis, selected crystals of as-prepared and heat-treated FeSe${}_{0.945}$Te${}_{0.055}$ were ground into powders. The XRD patterns of powders were collected using Rigaku D/Max RC diffractometer (Bragg-Brentano geometry, CuK$\alpha$ radiation, graphite analyzer crystal, scintillation counter). Full-profile calculations and phase quantitation were performed using the Rietveld method \cite{rietveld1969profile, bish1988quantitative} with derivative difference minimization routine implemented in DDM 1.95 software \cite{solovyov2004full}. The preferred orientation effect was taken into account within the framework of the March-Dollase model \cite{dollase1986correction}.

Magnetization was measured using a Quantum Design MPMS SQUID in a field of 10 Oe under zero-field cooled conditions and in field of 10 kOe for $\chi (T)$ dependence in the temperature range 10 - 300~K . 
Electrical measurements were done on cleaved samples with contacts  made by sputtering of Au/Ti layers.

\section{Results}
\begin{figure}[h]
\centering
  \includegraphics[scale=0.25]{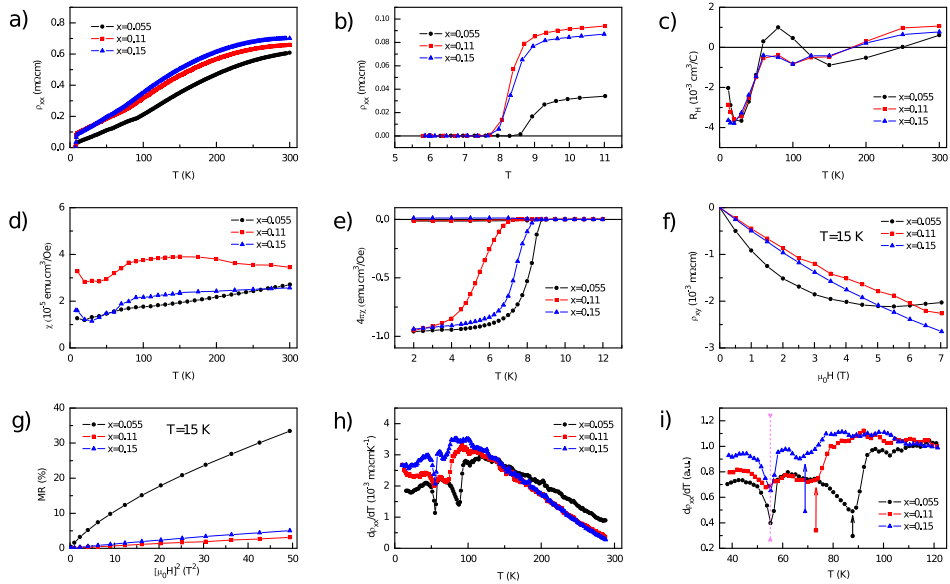}
  \caption{ The properties of as-prepared crystals of FeSe${}_{1-x}$Te${}_{x}$ with $x$=0.055, 0.11, and 0.15. (a)~Temperature dependence of the resistivity $\rho_{xx}$. (b)~Temperature dependence of the resistivity $\rho_{xx}$ at low temperatures. (c)~Temperature dependence of the Hall constant $R_{H}$. (d)~Temperature dependencies of the magnetic susceptibility $\chi$ in an applied field of 10 kOe. (e)~ZFC and FC magnetic susceptibility $\chi$ in an applied field of 30 Oe. (f)~Magnetic field dependence of the Hall resistivity $\rho_{xy}$. (g)~Magnetoresistance $MR$=($\rho_{xx}$(B)-$\rho_{xx}$(0))/$\rho_{xx}$(0) versus $B^{2}$ at 15~K. (h)~Temperature dependence of the derivative of the resistivity  $d\rho_{xx}/dT$. (i)~Temperature dependence of the derivative of the resistivity  $d\rho_{xx}/dT$ between 40~K and 120~K.}
  \label{fgr:fig1}
\end{figure}

The results of studying the transport properties of as-prepared crystals are shown in Fig.\ref{fgr:fig1}. The temperature dependence of the resistivity $\rho_{xx}$ has a similar metallic behavior for all compositions (Fig.\ref{fgr:fig1}, a). The temperatures of the superconducting transitions decrease insignificantly with increasing tellurium content (Fig.\ref{fgr:fig1}, b). The temperature dependence of the Hall constant also shows a slight change (Fig.\ref{fgr:fig1}, c). In particular, only one inversion point remains on $Rh(T)$ for $x=0.11$ and $x=0.15$.

The temperature dependencies of the magnetic susceptibility for compositions $x=0.11$ and $x=0.15$ have humps in the temperature range 100-200~K (Fig.\ref{fgr:fig1}, d), which may indicate the presence of a hexagonal phase. This is most likely because the synthesis temperature of these compositions was higher than the stability limit of the hexagonal phase for FeSe. Nevertheless, based on the values of the susceptibility of the hexagonal phase, the amplitudes of the humps indicate that the content of the hexagonal phase is negligible \cite{ovchenkov2019nematic}. Thus, we can assume that the susceptibility data indicate a high quality of the synthesized crystals.

The ZFC-FC curves for the samples composed of several plate crystals oriented parallel to the magnetic field (Fig.\ref{fgr:fig1}, e), show a full Meissner effect, although the systematic difference between the transition temperatures determined from magnetic and transport measurements may indicate a complex microstructure of crystals.

Similar to FeSe, the field dependencies of the Hall component of resistivity $\rho_{xy}$ have a nonlinear form at low temperatures, which indicates the presence of mobile carriers  \cite{SUST-30-3-035017}. As can be seen from Fig.\ref{fgr:fig1} f), the nonlinearity is noticeably suppressed with increasing tellurium content, which agrees with a significant decrease in the carrier mobility in these compositions.

The field dependencies of the magnetoresistance $MR$ (Fig.\ref{fgr:fig1} g) confirm a significant change in the carrier mobility. The slope of the $MR(B^{2})$ dependencies for the compounds under study, with fairly good accuracy, is equal to the square of the carrier mobility \cite{2017ovchenkovMISM}. Thus, a change in the magnitude of the magnetoresistance $MR$ in a field of 7 T from 33\% to 5\% corresponds to a change in the mobility by a factor of two and a half. This means that the studied compositions are close to the crossover from bad to good metal, which occurs in the 11 and some other series of IBS \cite{PhysRevB.88.094501}.

The resistivity of the samples is also sensitive to the transition from the high-temperature tetragonal phase to the low-temperature orthorhombic phase. The temperature dependence of the derivative of the resistivity  $d\rho_{xx}/dT$ exhibits anomalies in the region of the structural transition (Fig.\ref{fgr:fig1} h and \ref{fgr:fig1} i). The appearance of these anomalies makes it possible, in particular, to control the phase separation in the samples. It is known that phase separation in FeSe${}_{1-x}$Te${}_{x}$ with a low tellurium content occurs into two tetragonal phases with different lattice parameters  \cite{zhuang2014}. As seen from Fig.\ref{fgr:fig1} i), the anomaly for the studied compositions has a split form. This shows the presence of two distinct phases with different temperatures of the structural transition. It can be noted that while the position of the anomaly at higher temperatures is different for the samples, the position of the low-temperature anomaly is the same for all samples.

\begin{figure}[h]
\centering
  \includegraphics[scale=0.5]{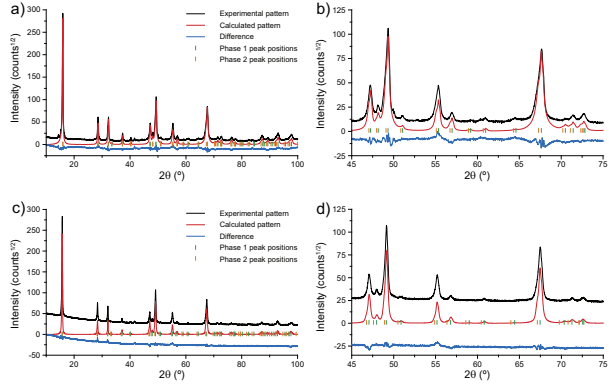}
  \caption{Full-profile analysis of diffraction patterns of as-prepared FeSe${}_{0.945}$Te${}_{0.055}$ (panels (a) and (b)) and heat-treated FeSe${}_{0.945}$Te${}_{0.055}$ (panels (c) and (d) for HT-A). Panels (b) and (d) are enlarged portions of (a) and (c), respectively. Intensity is given on a square root scale to enhance low-intensity peaks details. Peak positions of phases \#1 and \#2 are given as ticks.} 
  \label{fgr:fig_xrd}
\end{figure}

The analysis of XRD data for as-prepared FeSe${}_{0.945}$Te${}_{0.055}$ (Fig.\ref{fgr:fig_xrd}(a) and (b)) revealed that the main phase is the tetragonal polymorph of the FeSe structure \cite{hagg1933}. The significant asymmetry of the diffraction peaks is associated with the presence of a second isostructural phase with slightly different cell parameters. Its presence was confirmed by full-prole calculations. During the refinement, the parameters of the unit cells were calculated, as well mass fractions of the phases (Table \ref{tbl:T1}). During the analysis of XRD patterns of annealed samples (Fig.\ref{fgr:fig_xrd}(c) and (d) for HT-A), it was found that the content of the second phase decreases significantly. At the same time, the cell parameters are not changing significantly.

\begin{table*}[h]
\caption{\label{tbl:T1}  The results of the XRD analysis for as-prepared and heat-treated FeSe${}_{0.945}$Te${}_{0.055}$. The error in the content value ranges from about 1.5 to about 3\%. }
\lineup
\begin{tabular}{ccccc}
\br
Sample&Phase&Content (\%)& a (\r{A}) & c (\r{A})\\
\mr
as-prepared&\#1&52.3&3.7731(6)&5.5338(6)\\
as-prepared&\#2&47.7&3.787(2)&5.558(1)\\
\mr
HT-A&\#1&87.2&3.7725(9)&5.5334(3)\\
HT-A&\#2&12.8&3.799(5)&5.556(1)\\
\mr
HT-Q&\#1&84.3&3.773(1)&5.5351(4)\\
HT-Q&\#2&15.7&3.797(7)&5.557(1)\\
\br
\end{tabular}
\end{table*}

The crystals of FeSe${}_{1-x}$Te${}_{x}$ are often non-uniform at a microscopic level \cite{PROKES2015}. The temperature of synthesis is usually reduced at low tellurium content because of the shift of the stability boundary of the hexagonal phase. With a decrease in the synthesis temperature, a deterioration in the homogeneity of the tellurium distribution can be expected. We initially assumed that the phase separation at low $x$ can be due to the non-uniform distribution of tellurium and the formation of clusters with high tellurium content. We were going to find the optimal heat treatments that would allow achieving a more uniform distribution of tellurium. We expected to see the difference introduced by quenching, and we expected that long heat treatment at intermediate temperature should enhance phase separation. However, the results obtained do not confirm the original assumptions.

\begin{figure}[h]
\centering
  \includegraphics[scale=0.25]{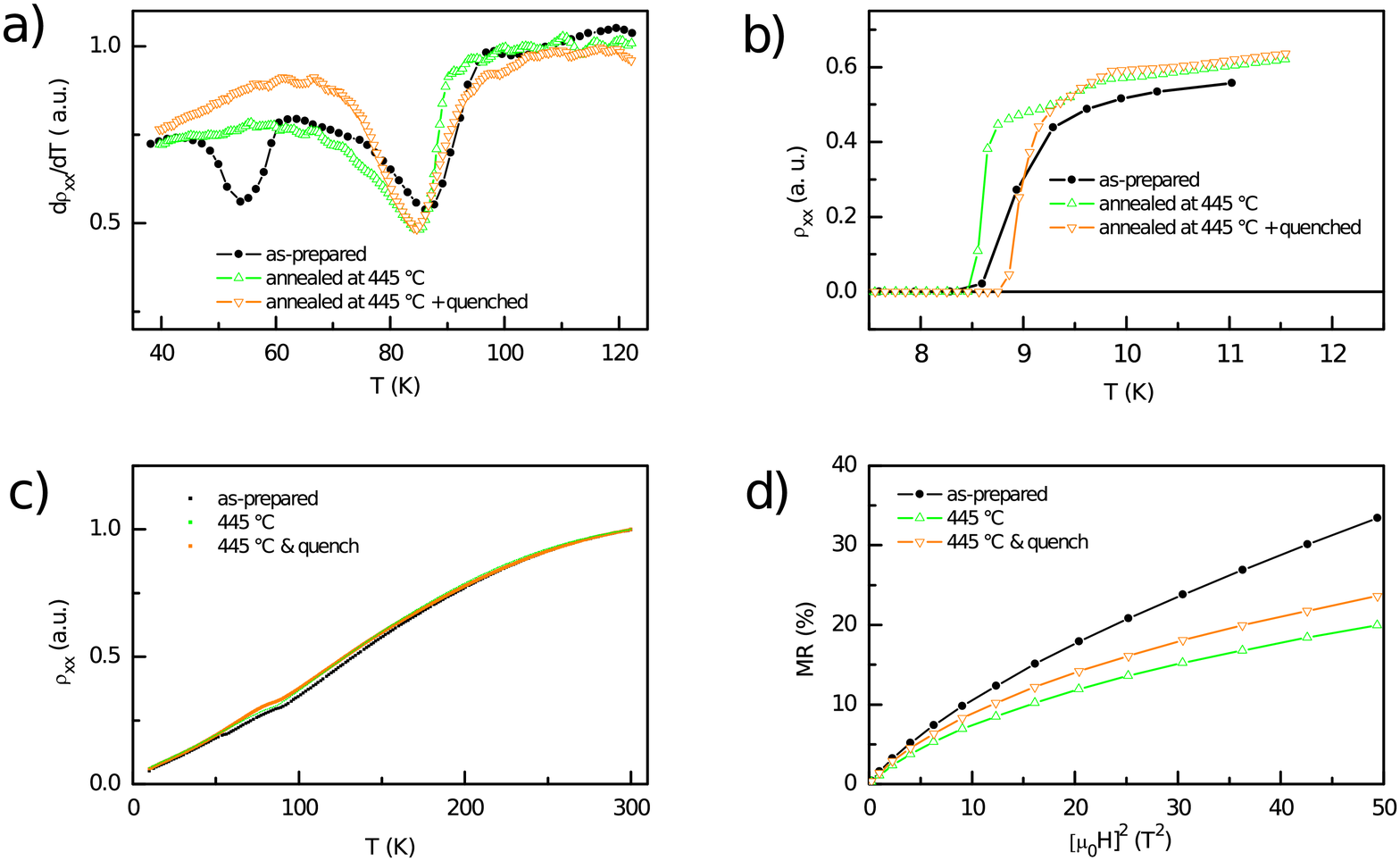}
  \caption{The properties of as-prepared and heat-treated crystals of FeSe${}_{0.945}$Te${}_{0.055}$. (a)~Temperature dependence of the derivative of the resistivity  $d\rho_{xx}/dT$ between 40~K and 120~K.  (b)~Temperature dependence of the resistivity $\rho_{xx}$ at low temperatures. (c)~Temperature dependence of the resistivity $\rho_{xx}$ normalized at 300~K. (d)~Magnetoresistance $MR$=($\rho_{xx}$(B)-$\rho_{xx}$(0))/$\rho_{xx}$(0) versus $B^{2}$ at 15~K. }
  \label{fgr:fig2}
\end{figure}

We found that heat treatment at a temperature close to the synthesis temperature already suppresses the second phase quite effectively. Fig.\ref{fgr:fig2} a) shows the $d\rho_{xx}/dT$ curves for the as-prepared crystal and two crystals heat-treated at 445~$^{\circ}$C. Both heat-treated samples have no low-temperature anomaly on the $d\rho_{xx}/dT$, although quenching changes the shape of $\rho_{xx}(T)$ near phase transitions, as can be seen from Fig.\ref{fgr:fig2} a) and  Fig.\ref{fgr:fig2} c).  In general, heat treatment had a rather slight effect on $\rho_{xx}(T)$ (Fig.\ref{fgr:fig2} c). And at the same time, the decrease in mobility at low temperatures by about 20-25\% (see the values of MR plotted in Fig.\ref{fgr:fig2} e), indicates a possible degradation of the microstructure of the sample in both heat treatments.

\begin{figure}[h]
\centering
  \includegraphics[scale=0.5]{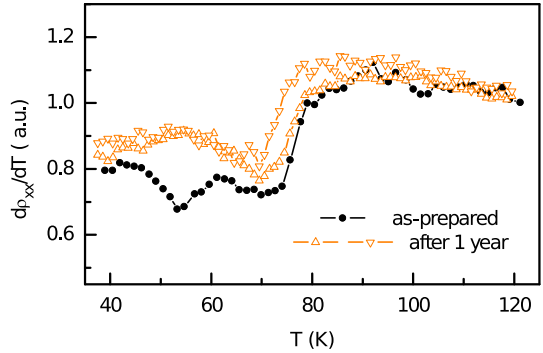}
  \caption{Temperature dependence of the derivative of the resistivity  $d\rho_{xx}/dT$ between 40~K and 120~K for FeSe${}_{0.89}$Te${}_{0.11}$ sample prepared immediately after synthesis and two samples prepared after one year of storage in an evacuated quartz ampoule.}
  \label{fgr:fig3}
\end{figure}

After heat treatment at 445~$^{\circ}$C, we carried out heat treatment at 150~$^{\circ}$C for two weeks. Our measurements did not reveal any changes in transport properties after the second heat treatment. However, we found that the phase composition can change during long-term storage at room temperature. For example, Fig.\ref{fgr:fig3} shows the $d\rho_{xx}/dT$ for the sample FeSe${}_{1-x}$Te${}_{x}$ with $x = 0.11$ prepared immediately after synthesis and two samples studied after long-term storage in an evacuated quartz ampoule.

Thus, the results obtained indicate that the main effect on phase separation could have been partial oxidation of the sample, which, as is well established  \cite{Sun_2019}, effectively removes excess iron. On the other hand, variations in the stoichiometry of iron are possible in all compositions of 11 series. Therefore, there must be some additional factor causing phase separation in the investigated range of compositions. Below, we will discuss one of the possible reasons for the structural instability of the iron plane for the FeSe${}_{1-x}$Te${}_{x}$ compositions with low tellurium content.

\section{Discussion}

For the iron-based superconductors, several types of electronic instability have been identified that may be related to  superconductivity in these compounds \cite{PhysRevLett.101.057003}. Structural instabilities can also be expected at certain values of the lattice parameters, which should be investigated in coordinate space. An effective way to consider the properties of compounds in real space is the molecular orbital (MO) method.

In the iron plane of IBS, there is a direct $d$-$d$ exchange between neighboring iron atoms, which can be considered separately from iron-chalcogen or iron-pnictogen bonds. The degree of participation of the direct $d$-$d$ exchange in the total energy of the chemical bond is measured by the dispersion of the corresponding bands, which is usually 3-4 eV for the IBS.

A simple illustrative model of the direct $d$-$d$ exchange is the Bethe-Slater curve, which describes the change in the sign of the exchange between neighboring $d$ ions depending on the distance between them. In body-centered cubic Fe the exchange has different sign for $e_{g}$ and $t_{2g}$ orbitals \cite{PhysRevLett.116.217202} and is strongly negative for the latter. In FeSe${}_{1-x}$Te${}_{x}$ the Fe-Fe distance is in the range 2.6-2.7~{\AA} \cite{ivanov2016local}, which is slightly larger than the value 2.5~{\AA} for the body-centered cubic Fe. Thus, the plane of iron atoms in  FeSe${}_{1-x}$Te${}_{x}$ can be close to the condition of the change in the sign of exchange for some orbitals.

A change in the sign of the exchange means some equilibrium between tensile and compressive strain for $d$ orbitals in the plane of iron atoms. At this point, the triplet and singlet $d$ orbitals are degenerate, which should lead to a significant rearrangement of the electronic structure near this point. This must be the point of instability for the electronic structure. We consider that a change in the sign of the exchange can account for the quantum criticality observed in FeSe \cite{PhysRevB.97.201102}, the anomalous behavior of FeSe in ultrasonic experiments \cite{epl_0295-5075-101-5-56005, fil2013piezomagnetism}, and the large elastoresistance effect \cite{Chu710}.

The low polarity of the iron-pnictogen bond may be an important component of electronic instability in 11 series. Low values of the Coulomb contribution to the energy of the crystal field, as well as close to zero values of the Madelung energy, facilitate the charge redistribution, which should occur during changes in the electronic structure. In our opinion, the inversion of the majority carriers is a consequence of these changes in the electronic structure near the point of inversion of the direct $d$-$d$ exchange.

The crystals FeSe$ _ {1-x}$Te$ _ {x}$ with a high tellurium content usually grow with an excess of iron while the crystals FeSe$ _ {1-x}$S$ _ {x}$ usually have an iron deficiency. Thus we can formally assume that the energy corresponding to the addition of one iron atom to the structure changes its sign along 11 series. Then there must be an equilibrium point for which adding iron to the structure costs zero energy. In our opinion, near this point, phases with different values of iron stoichiometry can exist simultaneously.

\section{Conclusion}

Materials with electronic instability can be of interest for various fields of application, for example, for piezoresistive and thermoelectric applications. It is interesting to consider ways to enhance the orbital degeneracy effect and electronic instability in FeSe. The substitution of elements can have limited success because the introduced disorder removes the degeneracy and increases the crystal field. Optimal doping probably can be achieved by the addition of intercalating layers, which can also suppress structural instability.

\section{Acknowledgments}

This work has been supported by Russian Foundation for Basic Research through Grant 20-02-00561A. OSV acknowledge the financial support by Russian Science Foundation through the project 19-42-02010. ESK and ANV  acknowledge the financial support by the Megagrant 2020-220-08-6358 of the Government Council on Grants of the Russian Federation.

\section*{References}
 \bibliographystyle{unsrt} 
\bibliography{FeSeTe_Sep.bib}


\listoffigures

\end{document}